# Effect of traps on the current impulse from X-ray induced conductivity in wide-gap semiconductors

V.Ya. Degoda[a], A.O. Sofiienko[b*]

[a] Taras Shevchenko National University of Kyiv, 64 Volodymyrs'ka, 01601 Kyiv, Ukraine,
degoda@univ.kiev.ua.
[b*] University of Bergen, 55 Allegaten, PO Box 7803, 5020 Bergen, Norway, off. 352,
asofienko@gmail.com

**Abstract.** This article presents a theoretical model for the calculation of the current impulse from X-ray induced conductivity in wide-gap semiconductors that contain different types of traps and recombination centres. The absorption of one X-ray photon in a semiconductor with ohmic contacts was investigated. The influence of the main parameters of the traps and recombination centres on the shape and amplitude of the current impulse was determined.

## 1. INTRODUCTION

Currently, semiconductor detectors for X-ray and gamma radiation are the most advanced devices for radiation monitoring in all industry branches, such as power engineering and science [1-4]. These detectors have a significant advantage over scintillation and gaseous detectors and have a wider potential for improvement and modernisation. One of the most advanced trends in semiconductor physics is the production and investigation of wide-gap semiconductor materials (e.g., ZnSe [5, 6] or SiC [7, 8]) that have a dielectric band gap ($E_g > 2.5$ eV). For moderate values of impurity concentrations and crystalline lattice flaws, these materials can be efficiently used to solve problems in radiometry and spectrometry over a wide temperature range [6, 9]. To use wide-gap semiconductors as radiation detectors in practical applications, the kinetic theory of X-ray conductivity needs to be developed. This theory will provide an understanding of the influence of shallow traps, deep traps and recombination centres on the shape and amplitude of the current pulse in a crystal after the absorption of one X-ray photon and will indicate when the charge collection efficiency will be much less than 100% due to these effects. A complete theoretical analysis of the influence of several types of semiconductor traps and recombination centres on current pulses that are generated by the absorption of X-rays has not been reported in the literature due to the high complexity of solving the charge transport equations for such a multiparametric task. Several simplified models have been presented and can be useful for understanding the influence of deep traps on the detector charge collection efficiency [10, 12]; however, these models only hold for a limited temperature range due to the dependence of the effective value of the trap localisation time on the crystal temperature [13]. Considering the above, a modern theoretical model of the X-ray induced conductivity of wide-gap semiconductors should take into account any type and number of traps or recombination centres to give the best agreement with real materials. For the theoretical consideration of the formation of current pulses from X-ray induced conductivity in wide-gap semiconductors, it is necessary to define the kinetics of the spatial distribution of the generated free electrons and holes, which are significantly influenced by different point defects in the crystal.

## 2. CURRENT PULSE OF AN IDEAL SEMICONDUCTOR FROM X-RAY EXCITATION

To develop a kinetic model of X-ray induced conductivity (XRC), it is first necessary to consider the kinetics of generated free charge carriers in the semiconductor after the absorption of one X-ray photon. Such kinetics is described by a generalised system of kinetic equations [14], which comprises two differential equations for free electrons and free holes, the Poisson equation for the spatial distribution of the electric field intensity in the semiconductor and equations for local centres, whose quantity corresponds to the number of types of local centres (shallow and deep traps for electrons and holes, as well as various types of recombination centres). It is impossible to solve such a system of differential equations and to simultaneously obtain the analytical dependences of the concentrations of free and localised charge carriers and charge-exchange recombination centres. Thus, the following approach is offered: the successive investigation of the problem from a model of an ideal semiconductor material to a real semiconductor material by adding new parameters step-by-step for the traps and recombination centres. The consideration of an ideal crystal is a convenient method for the development of a basic X-ray conductivity model. This model is further developed by adding in equations that describe the localisation of electrons and holes on the traps and their recombination as well as summands in the transport equation of free charge carriers. Such a successive development was partially shown in the works [15, 16] and enabled the determination of the influence of the main material characteristics on the charge collection kinetics in the crystal and on the X-ray conductivity value. The analysis for an ideal semiconductor (in which the system of kinetic equations is simplified to two equations for free electrons and holes) enables the correct calculation of the charge collection kinetics in a single crystal after the absorption of one X-ray photon. It is necessary to calculate the spatial distribution of the generated electrons and holes, $N^{\pm}(x, y, z, t)$, whose motion is determined by diffusion and drift in the electric field (if the semiconductor detector has ohmic contacts and the electric field intensity in the crystal is uniform).

$$N^{\pm}(x,y,z,t) = \frac{N_0}{4\pi D^{\pm} t} exp\left[-\frac{(y-y_0)^2 + (z-z_0)^2}{4D^{\pm} t} \pm \frac{\mu^{\pm} E_0 (x-x_0)}{2D^{\pm}} - \frac{(\mu^{\pm} E_0)^2 t}{4D^{\pm}}\right]$$

$$\times \frac{2}{d}\sum_{n=1}^{\infty}\left[exp\left(-\left(\frac{\pi n}{d}\right)^2 D^{\pm} t\right) sin\left(\frac{\pi n x}{d}\right) sin\left(\frac{\pi n x_0}{d}\right)\right], \qquad (1)$$

where $N_0$ is the number of generated electrons and holes after the absorption of one X-ray photon ($N_0 = h\nu_x/3E_g$), $x_0$ is the coordinate of the absorption, $d$ is the distance between the electrodes on the crystal, $E_0$ is the electric field intensity in the crystal (uniform field), $D^{\pm}$ is the diffusion constants of the charge carriers and $\mu^{\pm}$ is the mobility. The electrons and holes drop out of the general carrier distribution (1) when they reach the electrodes, which is why formula (1) satisfies the following boundary conditions: $N^{\pm}(0,y, z, t)=N^{\pm}(d, y,z, t)=0$. These conditions make it possible to obtain the correct normalisation for the carrier concentration, which decreases during drift in the crystal.

An analysis of the absorption process of one X-ray photon and of the succeeding relaxation of non-equilibrium electrons and holes shows that their proper electric field intensity rapidly relaxes for 0.1-1 ns in different materials [15], which means that the Coulomb interaction of non-equilibrium electrons and holes may be neglected. The drift current of the charge carriers ($q^-$ and $q^+$) in the external electric circuit is determined by the Shockley-Ramo relation [17]:

$$i(t) = \frac{E_0}{d}\left(q^-(t)\mu^- + q^+(t)\mu^+\right) = \frac{eN_0 E_0}{d}\left(E(t)\mu^- + P(t)\mu^+\right), \quad (2)$$

where "$e$" is the elementary charge and the functions $E(t)$ and $P(t)$ are relative quantities of the generated electrons and holes that remain free at some time "$t$" and that continue to drift towards the electrodes of the crystal:

$$E(t) = \frac{1}{N_0}\int_{-\infty}^{\infty}\int_{-\infty}^{\infty}\int_0^d N^-(x,y,z,t)\,dxdydz$$

$$P(t) = \frac{1}{N_0}\int_{-\infty}^{\infty}\int_{-\infty}^{\infty}\int_0^d N^+(x,y,z,t)\,dxdydz \quad (3)$$

The calculation of formulae (3) is significantly complicated by the existence of the infinite series in the accurate solution of the carrier concentration function (1). A detailed analysis of the accurate solution to formulae (3) enables the proposal of a simple analytic dependence for their approximation (if the electrical field is uniform in the crystal):

$$E(t) = \left[1 + \exp\left(\frac{\mu^- E_0 t - x_0}{\Delta^-(t)}\right)\right]^{-1}; \quad P(t) = \left[1 + \exp\left(\frac{\mu^+ E_0 t - (d - x_0)}{\Delta^+(t)}\right)\right]^{-1}, \quad (4)$$

where the function $\Delta^\pm(t)$ is responsible for the diffusive "blurring" of the spatial distribution of the charge carriers when they approach the detector electrodes. This effect directly depends on the diffusion coefficients of the carriers:

$$\Delta^\pm(t) = \frac{1}{2}\sqrt{2D^\pm t}. \quad (5)$$

The factor «½» is an approximation parameter for function (4), and $2D^\pm t$ is the dispersion of the spatial distribution of the electrons and holes due to their diffusion in the crystal. A detailed analysis of the approximate formulae (4) was performed in one of our previous papers [16]. The approximate formulae (4) and (5) simplify the calculation of the current pulse function:

$$i(t) = \frac{eN_0 E_0}{d}\left[\frac{\mu^-}{1 + \exp\left(2\frac{\mu^- E_0 t - x_0}{\sqrt{2D^- t}}\right)} + \frac{\mu^+}{1 + \exp\left(2\frac{\mu^+ E_0 t - (d - x_0)}{\sqrt{2D^+ t}}\right)}\right], \quad (6)$$

This formula keeps the dependence on the basic kinetic parameters of the carrier motion in an ideal semiconductor material. The calculations of current pulses upon the absorption of one X-ray photon (15 keV) at different places in the ideal semiconductor detectors Si (a) and ZnSe (b) are shown in Figure 1.

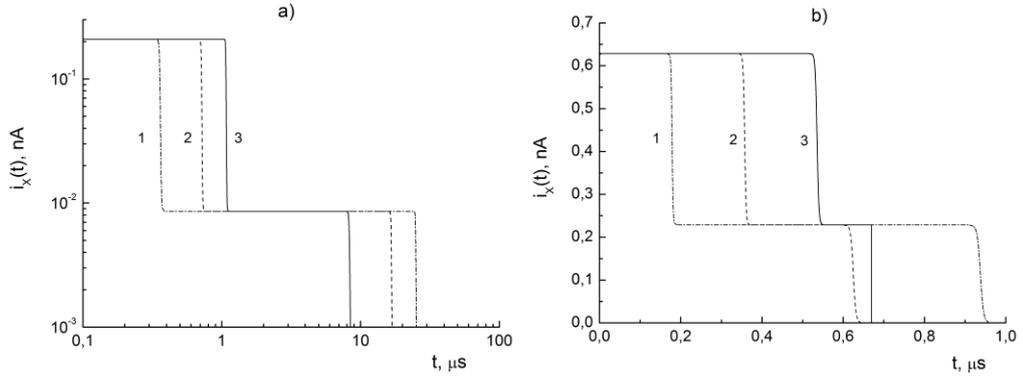

Fig. 1. The calculated current pulses in ZnSe (a) and Si (b) after the absorption of one X-ray photon (15 keV); $E_0 = 1000$ V/cm, d=1 cm, $N_0 = 1\,800$ (a) or $5\,000$ (b) and $x_0 = 0.25$ cm (1), 0.5 cm (2) or 0.75 cm (3).

Formulae (4) and (6) can be used to estimate the shape and amplitude of the current pulse in an ideal semiconductor, excluding the recombination processes. As will be shown below, these relations can be used with some changes when a semiconductor contains traps and recombination centres.

## 3. CURRENT PULSE IN A SEMICONDUCTOR WITH TRAPS AND RECOMBINATION CENTERS

### 3.1. Current pulse in a semiconductor with one type of shallow trap

We use the term "shallow traps" for traps that have a localisation time ($\tau_i$) for trapped electrons or holes that is less than or commensurable with the average drift time in the crystal. Additionally, it is assumed that the spatial distribution of shallow traps in the sample is homogeneous. For a semiconductor that only has shallow traps, the system of kinetic equations for free and trapped carriers becomes more complicated by the addition of two equations that describe trapping and delocalisation [14]. An exact solution of the system of kinetic equations can only be found using numerical methods. However, another method for finding the number of trapped and free carriers in the crystal can be proposed. If the lifetime of the charge carriers in the free state ($\tau_S^\pm$ is the average time between two successive trapping acts) is smaller than their drift time ($T_{dr}^\pm = \Delta x^\pm / (\mu^\pm \cdot E_0)$), which occurs for all wide-gap semiconductors with electrodes approximately 1 mm or more apart, then, after some time, a dynamic equilibrium will be reached between the number of free ($N^\pm$) and trapped ($N_0 - N^\pm$) charge carriers. Of course, for shallow traps, the number of filled traps compared with free ones can be neglected. This process is described by the balance equation that shows the relation between free and trapped carriers in the crystal without regard to their location (taking into account the drift and initial localisation of carriers $N_0$ that are generated in the crystal after the absorption of one X-ray photon):

$$\frac{dN^\pm}{dt} = -\frac{N^\pm}{\tau_S^\pm} + \frac{N_0 - N^\pm}{\tau_i^\pm} \Rightarrow N^\pm(t) = \frac{N_0}{\left(1 + \frac{\tau_i^\pm}{\tau_S^\pm}\right)}\left[1 + \frac{\tau_i^\pm}{\tau_S^\pm} e^{-t\left(\frac{1}{\tau_S^\pm} + \frac{1}{\tau_i^\pm}\right)}\right], \qquad (7)$$

Practically, formula (7) determines the number of drifting charge carriers that create an electric current in the external electric circuit, either for an infinite sample or for the initial stage of diffusion-drift motion $\left(t < T_{dr}^{\pm}\right)$ because the recombination of free carriers on the electrical contacts is disregarded. The relation $N^{\pm}(t)/N_0$ is the probability of each charge carrier being in a free state, independent of its spatial location in the crystal. The kinetics of the change in the free carrier number significantly depends not only on the type of shallow trap but also on the concentration, and this effect is shown in Figure 2. Formula (7) gives a statistically effective value $N^{\pm}(t)$ and does not take into account the fact that the full trapping time may be different for each particular electron or hole, as the number of trappings is a random value during drift.

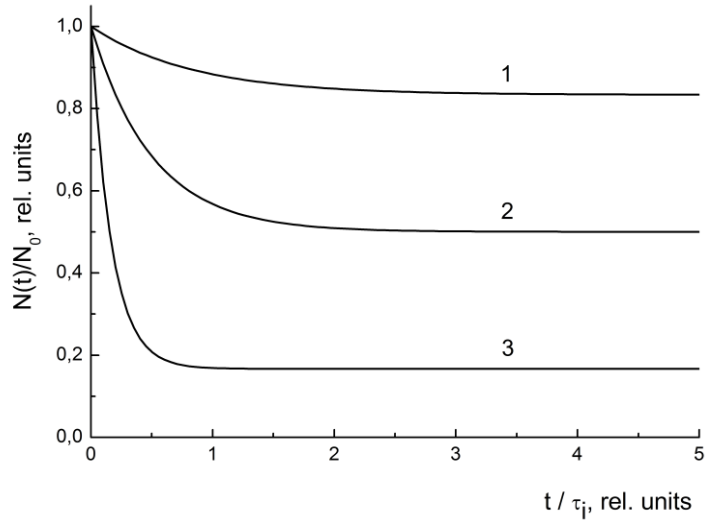

Fig. 2. The kinetics of the change in the relative number of free carriers ($N^{\pm}(t)/N_0$) in a semiconductor with different concentrations of shallow traps: $\tau_S = 5 \cdot \tau_i$ (1), $\tau_S = \tau_i$ (2) or $\tau_S = 0.2 \cdot \tau_i$ (3).

Obviously, the statistical character of the carrier trapping process should have some influence on the shape of the X-ray conductivity current impulse. The average lifetime of the charge carriers in their free state until the moment of trapping depends on the trap concentration ($v_i^{\pm}$), the cross-section of the trapping ($\sigma_i^{\pm}$) and the thermal velocity ($u^{\pm}$) in the following manner: $\tau_S^{\pm} = (v_i^{\pm}\sigma_i^{\pm}u^{\pm})^{-1}$ [13]. The average number of trappings ($m_0$) during drift can be defined by the relation $m_0^{\pm} = T_{dr}^{\pm}/\tau_S^{\pm}$. The average number of trappings is not necessarily a whole number because trapping is a statistically random process. It is obvious that the variable $m_0$ depends on the coordinate of the absorbed X-ray photon in the crystal ($x_0$):

$$\begin{cases} m_0^-(x_0) = \dfrac{T_{dr}^-}{\tau_S^-} = \dfrac{x_0 v_i^- \sigma_i^- u^-}{\mu^- E_0} \\ m_0^+(x_0) = \dfrac{T_{dr}^+}{\tau_S^+} = \dfrac{(d-x_0) v_i^+ \sigma_i^+ u^+}{\mu^+ E_0} \end{cases}. \qquad (8)$$

Taking into account the fact that the trapping acts are independent random processes, the probability of trapping can be written as a Poisson distribution: $F_m(m_0) = m_0^m e^{-m_0}/m!$. This equation enables the calculation of the number of carriers that were trapped exactly $m$-times during drift: $N_{0m} = N_0 \cdot F_m(m_0)$. The whole process of collecting $N_0$ carriers on the crystal electrode (for electrons or holes) can be divided into $m_{max}$ processes that are independent of each other:

$$m_{max}^{\pm} \approx m_0^{\pm} + 3\sqrt{m_0^{\pm}} \qquad (9)$$

Considering the gradual collection of charge on the crystal electrodes, it is necessary to take into account the diffusive character of the carrier motion during drift. Thus, formulae (4) are used with the following additions. For each $m$-group of carriers, the time of them staying in the detector will increase with increasing average trapping time on the traps: $\Delta T_m^{\pm} = m \cdot \tau_i^{\pm}$. The fact that the trapping time on each trap is a random value should be taken into account and means that the shape of the X-ray conductivity current impulse is influenced by the statistical character of the delocalisation process. The probability of being in a localised state is described by the exponential distribution $w_i^{\pm}(t) = \dfrac{1}{\tau_i^{\pm}} \exp\left(-\dfrac{t}{\tau_i^{\pm}}\right)$ [13], where, for trapping $m$-times, the total localisation time for each carrier is $\left(\Delta T_m^{\pm}\right)_\Sigma = \sum_m \left(\tau_i^{\pm}\right)^*$ (where the symbol "*" indicates that the trapping time on each trap is a random value). The sum of the random variables, which are distributed according to an exponential law, is also a random variable that follows a gamma distribution:

$$\Gamma^{\pm}\left(m, \tau_i^{\pm}, t\right) = \frac{t^{m-1}}{\left(\tau_i^{\pm}\right)^m \Gamma_E(m)} \exp\left(-\frac{t}{\tau_i^{\pm}}\right), \qquad (10)$$

where $\Gamma_E(m) = \int\limits_0^\infty p^{m-1} e^{-p} dp$ is Euler's gamma function. The dispersion of gamma distribution (10) is $DIS_{\Gamma m}^{\pm} = m \cdot \left(\tau_i^{\pm}\right)^2$. Taking into account the fact that the carriers drift towards the electrodes with velocity $\mu^{\pm} \cdot E_0$, it is possible to calculate the dispersion of the gamma distribution in units of "distance": $\left(DIS_{\Gamma m}^{\pm}\right)_L = m \cdot \left(\tau_i^{\pm} \mu^{\pm} E_0\right)^2$. Taking into account the statistics for the trapping time, formula (5) may depend not only on the dispersion value of the spatial distribution of the charge carriers but also on the dispersion of their time delay due to trapping: $\Delta_m^{\pm}(t) = \dfrac{1}{2}\sqrt{2D^{\pm} t + m\left(\tau_i^{\pm} \mu^{\pm} E_0\right)^2}$. Formulae (4) for the $m$-groups of carriers, which are localised on the traps for a time $\Delta T_m^{\pm}$ during drift, can be changed as follows:

$$E_m(t) = \left\{1 + \exp\left[2\frac{\mu^- E_0(t - m\tau_i^-) - x_0}{\sqrt{2D^- t + m\left(\tau_i^- \mu^- E_0\right)^2}}\right]\right\}^{-1}$$

$$P_m(t) = \left\{1 + \exp\left[2\frac{\mu^+ E_0(t - m\tau_i^+) - (d - x_0)}{\sqrt{2D^+ t + m\left(\tau_i^+ \mu^+ E_0\right)^2}}\right]\right\}^{-1} \qquad (11)$$

In these formulae, only the diffuse expansion and the time delay of the free carriers on shallow traps at the electrical contacts are considered. It should be remembered that the average number of trapping acts $\left(m_0^{\pm}\right)$ for electrons and holes may be different. Taking into account the different number of charge carriers that are trapped for different amounts of times and the fact that the composite probability for the carriers to be in the drift state is the sum of the products of the probabilities of the separate and independent localisation and diffusion-drift motion processes, the general functional connections $E(t)$ and $P(t)$ will have the following form:

$$\begin{cases} E(t) = \sum_m F_m(m_0^-) E_m(t) = \dfrac{\left\{1+\dfrac{\tau_i^-}{\tau_S^-}\exp\left[-t\left(\dfrac{1}{\tau_S^-}+\dfrac{1}{\tau_i^-}\right)\right]\right\}}{\left(1+\dfrac{\tau_i^-}{\tau_S^-}\right)} \sum_m \dfrac{m_0^m \exp(-m_0)}{m!\left\{1+\exp\left[2\dfrac{\mu^- E_0(t-m\tau_i^-)-x_0}{\sqrt{2D^- t + m\left(\tau_i^- \mu^- E_0\right)^2}}\right]\right\}} \\[2em] P(t) = \sum_m F_m(m_0^+) P_m(t) = \dfrac{\left\{1+\dfrac{\tau_i^+}{\tau_S^+}\exp\left[-t\left(\dfrac{1}{\tau_S^+}+\dfrac{1}{\tau_i^+}\right)\right]\right\}}{\left(1+\dfrac{\tau_i^+}{\tau_S^+}\right)} \sum_m \dfrac{m_0^m \exp(-m_0)}{m!\left\{1+\exp\left[2\dfrac{\mu^+ E_0(t-m\tau_i^+)-(d-x_0)}{\sqrt{2D^+ t + m\left(\tau_i^+ \mu^+ E_0\right)^2}}\right]\right\}} \end{cases} \quad (12)$$

The current impulse in the external electrical circuit is counted according to the Ramo-Shockley formula (2) using formula (12). In the limit of decreasing trap concentration, when $m_0^{\pm} \to 0$ ($\tau_S^{\pm}/T_{dr}^{\pm} \geq 1$), the current impulse asymptotically approaches the current impulse function (6) that conforms with the motion of carriers in an ideal crystal. The Poisson distribution is quite accurately approximated by a Gaussian normal distribution when $m_0^{\pm} > 10$, $F_m(m_0) = (2\pi m_0)^{-\frac{1}{2}} \exp\left[-(m-m_0)^2/2m_0\right]$, where $m_0$ is the distribution dispersion. This equation enables the use of integrals (with the variable "$m$") instead of sums in formula (12). This integration averages over all of the possible values of the trapping number "$m$" for the probability of carriers to be in the free state during drift. Because a Gaussian distribution is a symmetric function with respect to the mean observation "$m_0$", the result of the integration of the approximate functions $E(t)$ and $P(t)$ gives the following relation for the current impulse function:

$$i(t) = \dfrac{eN_0 E_0}{d}\left(\dfrac{\mu^-\left\{1+\dfrac{\tau_i^-}{\tau_S^-}\exp\left[-t\left(\dfrac{1}{\tau_S^-}+\dfrac{1}{\tau_i^-}\right)\right]\right\}}{\left(1+\dfrac{\tau_i^-}{\tau_S^-}\right)\left(1+\exp\left[2\dfrac{\mu^- E_0(t-m_0\tau_i^-)-x_0}{\sqrt{2D^- t + 2m_0\left(\tau_i^- \mu^- E_0\right)^2}}\right]\right)}\right)$$

$$+ \frac{\mu^+ \left\{ 1 + \frac{\tau_i^+}{\tau_S^+} \exp\left[ -t\left( \frac{1}{\tau_S^+} + \frac{1}{\tau_i^+} \right) \right] \right\}}{\left( 1 + \frac{\tau_i^+}{\tau_S^+} \right) \left\{ 1 + \exp\left[ 2 \frac{\mu^+ E_0(t - m_0^+ \tau_i^+) - (d - x_0)}{\sqrt{2D^+ t + 2m_0^+ \left( \tau_i^+ \mu^+ E_0 \right)^2}} \right] \right\}} \quad (13)$$

To calculate the current impulse in a crystal with one type of shallow trap for electrons and one type of shallow trap for holes, we only need to know the average number of carrier trappings $m_0^\pm(x_0)$ during the drift time, which is determined by the parameters of the traps, the coordinate of the absorption of the X-ray photon and the value of the electric field intensity in the crystal. The current pulses for Si and ZnSe with one type of shallow trap are shown in Figure 3.

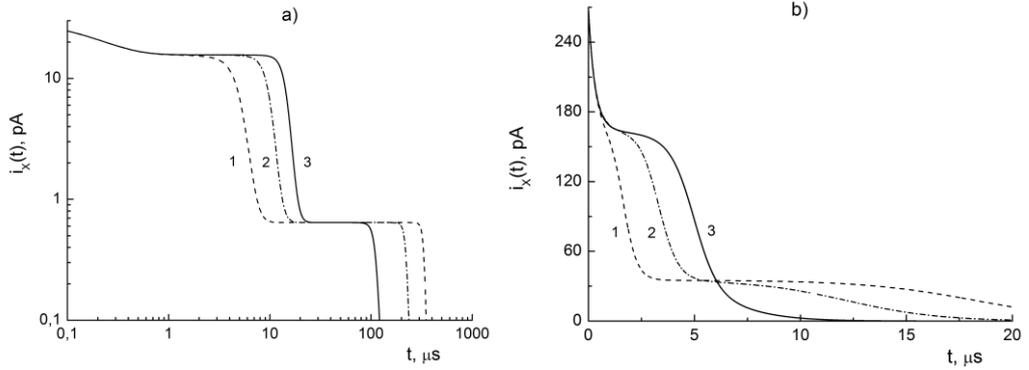

Fig. 3. Calculated current pulses in ZnSe (a) and Si (b) after the absorption of one X-ray photon (15 keV): $x_0$= 0.25 cm (1), 0.5 cm (2) or 0.75 cm (3); $E_0$ = 150 V/cm, d=1 cm, $N_0$ = 1 800 (a) or 5 000 (b), $v_i^+ = v_i^- = 10^{14}$ cm$^{-3}$, $\sigma_i^+ = \sigma_i^- = 10^{-15}$ cm$^2$, $\tau_i^- = 0.35$ μs and $\tau_i^+ = 1.0$ μs.

Thus, we have the time dependence of the X-ray conductivity current pulse in wide-gap semiconductors with one type of shallow trap.

### 3.2. Current impulse in a semiconductor with different types of shallow traps

Almost all real semiconductors have some concentrations of a few types of shallow traps for electrons $(v_i^-)$ and shallow traps for holes $(v_j^+)$, and the theoretical analysis should also be broadened for this case. We first need to determine how several types of traps in the crystal affect the diffusion-drift motion of free charge carriers. Because all trapping acts of carriers on shallow traps are independent processes, the character of their motion will not change, and the delay time will be defined as the additive sum of the trappings for all types of traps. It is logical to use formula (13) for the current pulse, replacing the parameters $\tau_S^\pm$, $\tau_i^\pm$, $m_0^\pm$ with the corresponding effective parameters $\tau_{S0}^\pm$, $\tau_{i0}^\pm$, $m_{00}^\pm$. Two main parameters have to be determined: the value of the average delays of the carriers during drift and the dispersion value of the distribution function of the trapping time on all types of traps. The average lifetime of electrons and holes in the free state will be determined by the total probability of trapping on all shallow traps:

$$\tau_{S0}^- = (u^- \sum_i \sigma_i^- v_i^-)^{-1}, \qquad \tau_{S0}^+ = (u^+ \sum_j \sigma_j^+ v_j^+)^{-1}. \qquad (14)$$

We use the facts that the traps are uniformly distributed in the sample and that their concentration ratio is constant. The average number of trappings during the drift time will depend on the absorption coordinates of the X-ray photon:

$$\begin{cases} m_{00}^-(x_0) = \sum_i m_{0i}^-(x_0) = \dfrac{x_0 u^-}{\mu^- E_0} \sum_i \left(v_i^- \sigma_i^-\right) \square\ x_0 \\ m_{00}^+(x_0) = \sum_j m_{0j}^+(x_0) = \dfrac{(d-x_0)u^+}{\mu^+ E_0} \sum_j \left(v_j^+ \sigma_j^+\right) \square\ (d-x_0) \end{cases}, \qquad (15)$$

The average trapping time of electrons and holes during drift will be the following:

$$\begin{cases} \Delta T_0^-(x_0) = \dfrac{x_0 \cdot u^-}{\mu^- \cdot E_0} \sum_i \left(v_i^- \sigma_i^- \tau_i^-\right) = m_{00}^- \cdot \tau_{i0}^- \Rightarrow \tau_{i0}^- = \dfrac{x_0 \cdot u^-}{m_{00}^-(x_0)\mu^- E_0} \sum_i \left(v_i^- \sigma_i^- \tau_i^-\right) = \dfrac{\sum_i \left(v_i^- \sigma_i^- \tau_i^-\right)}{\sum_i \left(v_i^- \sigma_i^-\right)} \\ \Delta T_0^+(x_0) = \dfrac{(d-x_0)u^+}{\mu^+ \cdot E_0} \sum_j \left(v_j^+ \sigma_j^+ \tau_j^+\right) = m_{00}^+ \cdot \tau_{j0}^+ \Rightarrow \tau_{j0}^+ = \dfrac{(d-x_0)\cdot u^+}{\mu^+ \cdot E_0} \sum_j \left(v_j^+ \sigma_j^+ \tau_j^+\right) = \dfrac{\sum_j \left(v_j^+ \sigma_j^+ \tau_j^+\right)}{\sum_j \left(v_j^+ \sigma_j^+\right)} \end{cases} \qquad (16)$$

There is a need to examine whether these effective parameters satisfy balance equation (7). The first term that determines the trapping of free carriers is the following:

$$N^\pm u^\pm \sum_i \sigma_i^\pm v_i^\pm = \dfrac{N^\pm}{\tau_{S0}^\pm}, \qquad (17)$$

and corresponds to the balance equation. The second term in the balance equation defines the thermal delocalisation of the carriers from traps:

$$\sum_i n_i^\pm w_i = \sum_i \dfrac{n_i^\pm}{\tau_i^\pm} = \sum_i \dfrac{(N_0 - N^\pm)}{\tau_i^\pm} \dfrac{\sigma_i^\pm v_i^\pm \tau_i^\pm}{\sum \sigma_i^\pm v_i^\pm \tau_i^\pm} = \left(N_0 - N^\pm\right) \dfrac{\sum_i \sigma_i^\pm v_i^\pm}{\sum_i \sigma_i^\pm v_i^\pm \tau_i^\pm} = \dfrac{N_0 - N^\pm}{\tau_{i0}^\pm}, \qquad (18)$$

This equation also corresponds to the balance equation if the number of trapped carriers on $i$-type traps, which is defined by the relative probability of trapping for some time in this state, is taken into account: $n_i^\pm = \left(N_0 - N^\pm\right) \dfrac{\sigma_i^\pm v_i^\pm \tau_i^\pm}{\sum_i \sigma_i^\pm v_i^\pm \tau_i^\pm}$. Thus, the effective parameters that are determined in this way, $\tau_{S0}^\pm$ i $\tau_{i0}^\pm$, fit the balance equation. The distribution function of the trapping time on traps of all types will be determined by the gamma distribution that takes into account the extended average trapping time, formula (16). The dispersion of the gamma distribution for the average number of trappings on traps of all types is:

$$DIS_{\Gamma m}^\pm = \sum_i m_0^\pm \cdot \left(\tau_i^\pm\right)^2 \approx \sum_i m_0^\pm \tau_i^\pm \tau_{i0}^\pm = m_{00}^\pm \left(\tau_{i0}^\pm\right)^2 \qquad (19)$$

Hence, when there are few types of shallow traps in the semiconductor material, the calculation of the current impulse can generally be performed using formula (15) with the effective parameters for the lifetime of free and localised states $\left(\tau_{S0}^\pm, \tau_{i0}^\pm\right)$ and the average total number of trapping acts $\left(m_{00}^\pm\right)$.

### 3.3. Current impulse in a semiconductor with deep traps and recombination centres

For a semiconductor that only has deep traps or recombination centres with some constant concentration ($v_D^{\pm}$), the transition of free charge carriers in the localised state becomes a principal process. Such a transition decreases the concentration of free carriers and the current impulse in the external circuit. We use the term "deep traps" for traps that have a trapping time that is much longer than the maximum time of the diffusion-drift motion $\left(\tau_i^{\pm} \gg d/\mu^{\pm} E_0\right)$ of the carriers to the electrical contacts. Because we investigate the kinetics for just one current impulse, such a term is valid for deep traps and for the recombination centres. We shall assume that the probability of trapping is the same for all carriers during the drift motion in all parts of the sample. Namely, $\sum_i u^{\pm} \sigma_i^{\pm} v_{iD}^{\pm} = Const(x,t)$, which means that the concentration of the unfilled deep traps and recombination centres is large enough for the trapping of all of the generated carriers and that the trap concentration practically cannot change. In this case, $\tau_D^{\pm}$ is the lifetime of a carrier in a free state, which will be same for all electrons and holes during the whole period of their diffusion-drift motion in the crystal:

$$\left(\tau_D^{\pm}\right)^{-1} = \sum_i u^{\pm} \sigma_i^{\pm} v_{iD}^{\pm} \qquad (20)$$

The kinetic equation for the free generated carriers after the absorption of one X-ray photon may be solved as:

$$\frac{dN^{\pm}}{dt} = -\frac{N^{\pm}}{\tau_D^{\pm}} \quad \Rightarrow \quad N^{\pm}(t) = N_0 e^{-\frac{t}{\tau_D^{\pm}}} \qquad (21)$$

Thus, the number of free carriers will exponentially decrease with time, and, in the case when the concentration of the deep traps is significant, all of the free carriers will reach the electrical contacts. According to the same law, the functions $E(t)$ and $P(t)$ will decrease, and free electrons and holes will move as in an ideal semiconductor until trapping or recombination (4). Thus, the current impulse in the external electric circuit can be determined using the following function:

$$i(t) = \frac{eN_0 E_0}{d} \left[ \frac{\mu^- \exp\left(-\frac{t}{\tau_D^-}\right)}{1 + \exp\left(2\frac{\mu^- E_0 t - x_0}{\sqrt{2D^- t}}\right)} + \frac{\mu^+ \exp\left(-\frac{t}{\tau_D^+}\right)}{1 + \exp\left(2\frac{\mu^+ E_0 t - (d - x_0)}{\sqrt{2D^+ t}}\right)} \right]. \qquad (22)$$

If $\tau_D^{\pm} \ll T_{dr}^{\pm}$, the equation will be simplified:

$$i(t) = \frac{eN_0 E_0}{d} \left[ \mu^- \exp\left(-\frac{t}{\tau_D^-}\right) + \mu^+ \exp\left(-\frac{t}{\tau_D^+}\right) \right]. \qquad (23)$$

The current impulses are the same regardless of the absorption coordinates of the X-ray photon ($x_0$). It should be noted that the amplitude of the current impulses is directly proportional to $N_0$, which is related to the X-ray photon energy. This fact is used very often for semiconductor radiation spectrometers. However, real semiconductors always have some deep traps and some shallow traps. Thus, it is necessary to define the X-ray conductivity current impulse for a sample that has all types of traps. In this case, in contrast to formula (7), the total number of carriers that create the current impulse in the external electric circuit gradually decreases due to their recombination or trapping in deep traps. The system of kinetic equations for free and trapped charge carriers is as follows:

$$\begin{cases} \dfrac{dN^{\pm}}{dt} = -\dfrac{N^{\pm}}{\tau_{S0}^{\pm}} + \dfrac{n_{S0}^{\pm}}{\tau_{i0}^{\pm}} - \dfrac{N^{\pm}}{\tau_{D}^{\pm}} \\ \dfrac{dn_{S0}^{\pm}}{dt} = \dfrac{N^{\pm}}{\tau_{S0}^{\pm}} - \dfrac{n_{S0}^{\pm}}{\tau_{i0}^{\pm}} \\ \dfrac{dn_{D}^{\pm}}{dt} = \dfrac{N^{\pm}}{\tau_{D}^{\pm}} \\ N_{0}^{\pm} = N^{\pm} + n_{S0}^{\pm} + n_{D}^{\pm} \end{cases} \quad , \qquad (24)$$

where $n_D^{\pm}$ is the number of filled deep traps for electrons or holes. The last equation in presented system (24) is the balance equation. The system of equations (24) does not have an analytical solution. However, to calculate the current impulse that is created by free electrons and holes, it is enough to obtain the solution as a function of the free charge carriers $N^{\pm} = f(t)$. The following original way can be proposed. First, we can use the balance equation to estimate the function of the filled shallow traps: $n_{S0}^{\pm}(t) = N_0^{\pm} - n_D^{\pm}(t) - N^{\pm}(t)$. The difference $\Delta N^{\pm}(t) = N_0^{\pm} - n_D^{\pm}(t)$ is the number of free or trapped charge carriers on shallow traps at any time. However, trappings on shallow and deep traps are independent random processes, and, thus, the function $\Delta N^{\pm}(t)$ is equal to the function of the free charge carriers when there are just deep traps in the crystal: $\Delta N^{\pm}(t) = N_0^{\pm} \cdot \exp\left(-t/\tau_D^{\pm}\right)$. This simple physical analysis makes it possible to estimate the function of the filled shallow traps ($n_{S0}^{\pm}(t)$) and to simplify the first kinetic equation in system (24):

$$\frac{dN^{\pm}}{dt} = -\frac{N^{\pm}}{\tau_{S0}^{\pm}} + \frac{N_0 e^{-\frac{t}{\tau_D^{\pm}}} - N^{\pm}}{\tau_{i0}^{\pm}} - \frac{N^{\pm}}{\tau_D^{\pm}} \qquad (25)$$

Kinetic equation (25) also does not have a simple analytical solution but can be solved in the following way. Because trapping on traps are independent processes, we can find the function $N^{\pm}$ as a product of two probabilities: the probability that the charge carriers are free after interaction with shallow traps and the probability that the charge carriers are free after interaction with deep traps, $N^{\pm}(t) = N_0 \cdot W_{shallow}(t) \cdot W_{deep}(t)$. Each of these probabilities can be found from the different kinetic equations (7) and (21). The final solution for the function of the free electrons and holes in the crystal is the following:

$$N^{\pm}(t) = \frac{N_0 e^{-\frac{t}{\tau_D^{\pm}}}}{\left(1 + \dfrac{\tau_{i0}^{\pm}}{\tau_{S0}^{\pm}}\right)} \left[1 + \frac{\tau_{i0}^{\pm}}{\tau_{S0}^{\pm}} e^{-t\left(\frac{1}{\tau_{S0}^{\pm}} + \frac{1}{\tau_{i0}^{\pm}}\right)}\right] \qquad (26)$$

To check the accuracy of the found function (26), the initial differential equation (25) was used. The obtained errors were less than $10^{-4}$% for any parameters of the shallow and deep traps. The current impulse is determined by:

$$i(t) = \frac{eN_0 E_0}{d} \left( \frac{\mu^- \left\{ 1 + \frac{\tau_{i0}^-}{\tau_{S0}^-} \exp\left[ -t\left( \frac{1}{\tau_{S0}^-} + \frac{1}{\tau_{i0}^-} \right) \right] \right\} \exp\left( -\frac{t}{\tau_D^-} \right)}{\left( 1 + \frac{\tau_{i0}^-}{\tau_{S0}^-} \right) \left\{ 1 + \exp\left[ 2 \frac{\mu^- E_0 (t - m_{00}^- \tau_{i0}^-) - x_0}{\sqrt{2D^- t + 2m_{00}^- \left( \tau_{i0}^- \mu^- E_0 \right)^2}} \right] \right\}} \right.$$

$$\left. + \frac{\mu^+ \left\{ 1 + \frac{\tau_{i0}^+}{\tau_{S0}^+} \exp\left[ -t\left( \frac{1}{\tau_{S0}^+} + \frac{1}{\tau_{i0}^+} \right) \right] \right\} \exp\left( -\frac{t}{\tau_D^+} \right)}{\left( 1 + \frac{\tau_{i0}^+}{\tau_{S0}^+} \right) \left\{ 1 + \exp\left[ 2 \frac{\mu^+ E_0 (t - m_{00}^+ \tau_{i0}^+) - (d - x_0)}{\sqrt{2D^+ t + 2m_{00}^+ \left( \tau_{i0}^+ \mu^+ E_0 \right)^2}} \right] \right\}} \right) \quad (27)$$

This obtained relation of the current impulse is the closest to the real model of a semiconductor and provides a logical solution for any proportion of trap concentrations because it takes into account the occurrence of shallow and deep traps and recombination centres in the material. The calculated current impulses using formula (26) are shown in Figure 4.

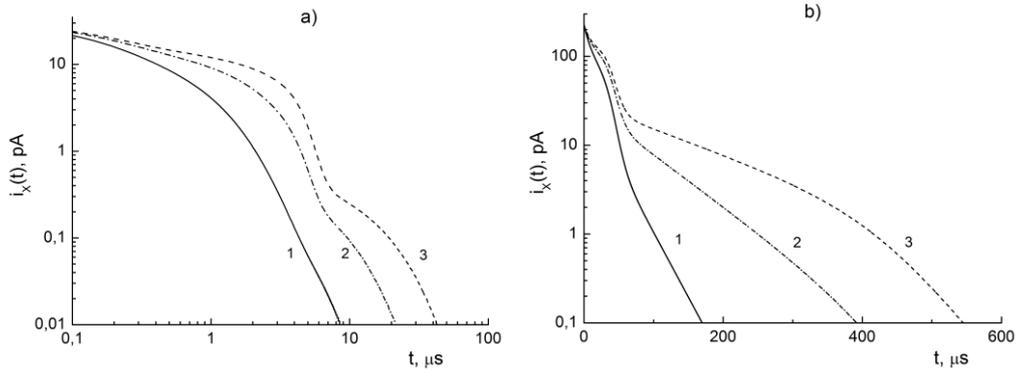

Fig. 4. The calculated current impulses after the absorption of one X-ray photon (15 keV) in ZnSe (a) and Si (b); $E_0 = 150$ V/cm, d=1 cm and $N_0 = 1\,800$ (a) or 5 000 (b) for different deep trap concentrations: $v_{iD}^\pm = 5\cdot 10^{13}$ cm$^{-3}$ (1), $v_{iD}^\pm = 2\cdot 10^{13}$ cm$^{-3}$ (2) or $v_{iD}^\pm = 1\cdot 10^{13}$ cm$^{-3}$ (3); $x_0 = 0.5$ cm, $v_i^+ = v_i^- = 10^{14}$ cm$^{-3}$, $\sigma_i^+ = \sigma_i^- = 10^{-15}$ cm$^2$, $\tau_i^- = 0.35$ μs and $\tau_i^+ = 1.0$ μs.

The conducted analysis can be supplemented by the case when the current impulse depends on deep traps that have a lifetime that is close to the total drift time of the carriers. The existence of such traps in a semiconductor will not have a significant effect on the shape of the current pulse but will reduce the carrier lifetime in the free state ($\tau_D^\pm$). Actually, the delocalisation of the carriers from such average and deep traps will create a small additional current in the semiconductor for quite a long time $\left( t \gg T_{dr}^\pm \right)$.

## 4. CONCLUSIONS

As a result of the sequential analysis of several theoretical models of a wide-gap semiconductor, from ideal to real, with shallow and deep traps or recombination centres, we obtained theoretical solutions for the current impulses in the external electric circuit of a semiconductor detector after the absorption of one X-ray photon. The case of ohmic contacts and a uniform electric field intensity in the crystal were investigated. A new probabilistic approach was used to determine simple analytical solutions for the current impulse in a semiconductor with shallow and deep traps or recombination centres. It was shown that the statistical character of the multiple trapping processes of the charge carriers on shallow and deep traps has a significant influence on the shape and amplitude of the X-ray conductivity current pulse. The obtained results can be useful for certain practical applications that need wide-gap semiconductors with large sizes and a wide operational temperature range when the value of the charge collection efficiency is considerably less than 100%.

## References

[1] G. Lutz, NIM Section A. **501** (2003) 288-297.
[2] A. Owens, A. Peacock, NIM Section A. **531** (2004) 18-37.
[3] A.V. Rybka, L.N. Davydova, I.N. Shlyakhova, V.E. Kutnya, I.M. Prokhoretza, D.V. Kutnya and A.N. Orobinsky, NIM Section A. **531** (2004) 147-156.
[4] P.J. Sellina and J. Vaitkus, NIM Section A. **557** (2006) 479-489.
[5] V.Ya. Degoda, A.O. Sofiienko, Semiconductors. **44** (2010) 1-7.
[6] A.O. Sofiienko, V.Ya. Degoda, Radiation Measurements. **47** (2012) 27-29.
[7] H.H. Jang, Y.K. Kima, S.H. Parka and S.M. Kang, NIM Section A. **579** (2007) 141-144.
[8] A.M. Ivanov, N.B. Strokan, A.A. Lebedev, NIM Section A. **597** (2008) 203-206.
[9] E. Kalinina, N. Strokan, A.M. Ivanov, A. Sadohin, A. Azarov, V. Kossov, R. Yafaev, S. Lashaev, Materials science forum, **556-557** (2007) 941-944.
[10] M.Bruzzi, D.Menichelli, S.Sciortino, J. Appl. Phys. **91**, 9 (2002) 5765-5774.
[11] M.J. Harrison, A. Kargar, D.S. McGregor, NIM Section A. **579** (2007) 134-137.
[12] J.D. Eskin, H.H. Barrett, H.B. Barber, J. of App. Phys., 85, 2 (1999) 647-659.
[13] V.V. Antonov-Romanovskii, Kinetics of the Photoluminescence of Crystal Phosphors [in Russian], Nauka, Moscow, 1966.
[14] V.Ya. Degoda, Vestn. KNU Shevchenko, Ser. Fiz. (special issue) **1** (2003) 47–52.
[15] V.Ya. Degoda, A.O. Sofiienko, Acta physica polonica A. **117**, 1 (2010) 333-338.
[16] V.Ya. Degoda, A.O. Sofiienko, Ukr. J. Phys. **55**, 2 (2010) 200-206.
[17] W. Shockley: J.Appl. Phys. **9** (1938) 635-638.